\documentclass[aps, pra, twocolumn, a4paper, groupedaddress, floatfix, 10pt, longbibliography]{revtex4-1}

\usepackage{graphicx, amsmath, amssymb, amsfonts, dsfont, enumitem, MnSymbol, comment, bm, color, hyperref}

\newcommand{\mf}[1]{\boldsymbol{#1}}
\newcommand{\ket}[1]{\ensuremath{|#1\rangle}}
\newcommand{\mc}[1]{\ensuremath{\mathcal{#1}}}

\newcommand{\braket}[2]{\ensuremath{\langle #1 | #2 \rangle}}

\newcommand{\len}{\ell}

\newcommand{\ma}[1]{\ensuremath{ \langle #1  \rangle}_{\text{anc}}}

\newcommand{\mcl}[1]{\ensuremath{ \llangle#1  \rrangle}}

\newcommand{\imag}{\mathrm{i}}

\begin{document}

\title{Variational quantum algorithms for nonlinear problems}

\author{Michael Lubasch${}^1$}

\author{Jaewoo Joo${}^1$}

\author{Pierre Moinier${}^2$}

\author{Martin Kiffner${}^{3,1}$}

\author{Dieter Jaksch${}^{1,3}$}

\affiliation{Clarendon Laboratory, University of Oxford, Parks Road, Oxford OX1 3PU, United Kingdom${}^1$}
\affiliation{BAE Systems, Computational Engineering, Buckingham House, FPC 267 PO Box 5, Filton, Bristol BS34 7QW, United Kingdom${}^2$}
\affiliation{Centre for Quantum Technologies, National University of Singapore, 3 Science Drive 2, Singapore 117543${}^3$}

\begin{abstract}
We show that nonlinear problems including nonlinear partial differential equations can be efficiently solved by variational quantum computing.
We achieve this by utilizing multiple copies of variational quantum states to treat nonlinearities efficiently and by introducing tensor networks as a programming paradigm.
The key concepts of the algorithm are demonstrated for the nonlinear Schr\"{o}dinger equation as a canonical example.
We numerically show that the variational quantum ansatz can be exponentially more efficient than matrix product states and present experimental proof-of-principle results obtained on an IBM Q device.
\end{abstract}

\maketitle

Nonlinear problems are ubiquitous in all fields of science and engineering and often appear in the form of nonlinear partial differential equations (PDEs).
Standard numerical approaches seek solutions to PDEs on discrete grids.
However, many problems of interest require extremely large grid sizes for achieving accurate results, in particular in the presence of unstable or chaotic behaviour that is typical for nonlinear problems~\cite{Wi03, AgLe09, St15}.
Examples include large-scale simulations for reliable weather forecasts~\cite{Ly06, Wa11, Pi13} and computational fluid dynamics~\cite{Mi73, FePe02, VeMa07}.

Quantum computers promise to solve problems that are intractable on conventional, i.e.\ standard classical, computers through their quantum-enhanced capabilities.
In the context of PDEs, it has been realized that quantum computers can solve the Schr\"{o}dinger equation faster than conventional computers~\cite{Ll96, AsEtAl05, KaEtAl08}, and these ideas have been generalized recently to other linear PDEs~\cite{HaHaLl09, Be14, BeEtAl15, MoPa16, KiEtAl17, ChKoSo17}.
However, nonlinear problems are intrinsically difficult to solve on a quantum computer due to the linear nature of the underlying framework of quantum mechanics.

Recently, the concept of variational quantum computing (VQC) attracted considerable interest~\cite{PeEtAl14, McRoBaAs16, MaEtAl16, KrClJa16, KrEtAl16, LiBe17, KaEtAl17, CoEtAl18, DuEtAl18, HeEtAl18, RoEtAl18, McEtAl18, EnLiBeYu18, ChEtAl19} for solving optimization problems.
VQC is a quantum-classical hybrid approach where the evaluation of the cost function ${\cal C}({\mf \lambda})$ is delegated to a quantum computer, while the optimization of variational parameters $\mf \lambda$ is performed on a conventional classical computer.
The concept of VQC has been applied, e.g., to simulating the dynamics of strongly correlated electrons through non-equilibrium dynamical mean field theory~\cite{GeEtAl96, KoEtAl06, KrClJa16, KrEtAl16}, and quantum chemistry calculations were successfully carried out on existing noisy superconducting~\cite{MaEtAl16, KaEtAl17, CoEtAl18} and ion quantum computers~\cite{HeEtAl18}.

\begin{figure}[t!]
\centering
\includegraphics[width=86.202mm]{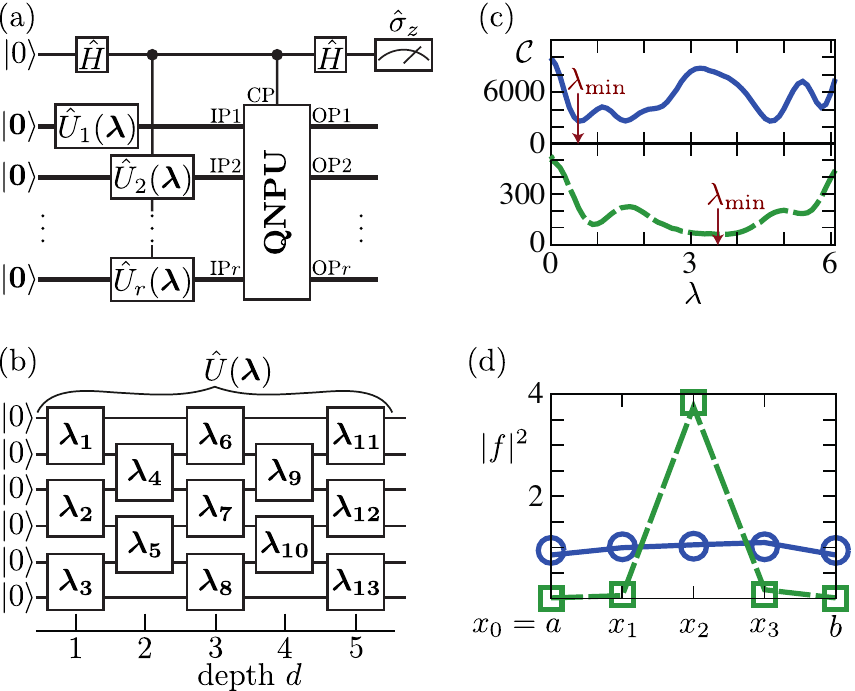}
\caption{\label{fig:1}
(a) Quantum network for summing a nonlinear function of the form $F = {f^{(1)}}^{*} \prod_{j=1}^{r} (O_{j} f^{(j)})$.
The ancilla qubit on the top line undergoes Hadamard gates $\hat{H}$ and controls operations of the rest of the network via the control port (CP) and is measured in the computational basis.
Starting from product states $\ket{\mf 0}$ of $n$ qubits shown as thick lines, variational states $\ket{\psi({\mf \lambda})_{j}} = \hat{U}_{j}({\mf \lambda})\ket{\mf 0}$ representing the functions $f^{(j)}$ are created and fed to the QNPU through input ports IP.
The QNPU contains problem specific quantum networks defining the linear operators $O_{j}$ and has the output ports OP.
(b) Network $\hat{U}({\mf \lambda})$ of depth $d = 5$ with $n = 6$.
The values ${\mf \lambda} = \{{\mf \lambda}_{1}, {\mf \lambda}_{2}, \ldots \}$ determine the form of the two-qubit gates.
(c) Cost function $\mc{C}(\lambda)$ for the nonlinear Schr\"{o}dinger equation Eq.~\eqref{gpe} for a harmonic potential $V$, a single variational parameter $\lambda$, and $g = 10^{4}$ (solid line) and $g = 10$ (dashed line).
The arrows indicate optimal values $\lambda = \lambda_{\rm min}$.
(d) Solutions $|f(x)|^{2}$ for $\lambda = \lambda_{\rm min}$ with $N = 4$ grid points and periodic boundary conditions $f(b) = f(a)$ for $g = 10^{4}$ (solid line) and $g = 10$ (dashed line).
The circles and squares are numerically exact results.
Experimental data in (c) and (d) was obtained -- from a modified circuit -- on IBM Q devices (see~\cite{SupplementalMaterial} for details).
}
\end{figure}

We extend and adapt the concept of VQC to solving nonlinear problems efficiently on a quantum computer by virtue of two key concepts.
First, we introduce a quantum nonlinear processing unit (QNPU) that efficiently calculates nonlinear functions of the form $F = {f^{(1)}}^{*} \prod_{j=1}^{r} (O_{j} f^{(j)})$ for VQC.
Measuring the ancilla qubit connected to the QNPU as shown in Fig.~\ref{fig:1}(a) directly yields the sum of all function values $\sum_{k} \Re \{ F_{k} \}$, where $\Re \{ \cdot \}$ denotes the real part.
The functions $f^{(j)}$ are encoded in variational $n$-qubit states $\ket{\psi({\mf \lambda})_j} = \hat{U}_{j}({\mf \lambda})\ket{\mf 0}$ created by networks of the form shown in Fig.~\ref{fig:1}(b).
The same function $f^{(i)}=f^{(j)}$ may appear multiple times by choosing $\hat{U}_{i}({\mf \lambda}) = \hat{U}_{j}({\mf \lambda})$.
Second, we use tensor networks as a programming paradigm for QNPUs to create optimized circuits that efficiently calculate linear operators $O_{j}$ acting on functions $f^{(j)}$.
In this way all quantum resources of nonlinear VQC scale polynomially with the number of qubits, which represents an exponential reduction compared with some conventional algorithms.

The variational states $\ket{\psi({\mf \lambda})_{j}}$ represent $N = 2^{n}$ values of the functions $f^{(j)}$ which form a trial solution to the problem of interest.
The cost function ${\cal C}({\mf \lambda})$ for nonlinear VQC is built up from outputs of different QNPUs that are then processed classically to iteratively determine the optimal set $\mf \lambda$.
Large grid sizes that are intractable on a conventional computer require only $n \gtrsim 20$ qubits which is within reach of noisy intermediate-scale quantum (NISQ) devices.
In addition, the scheme is applicable to other types of nonlinear problems that can be solved via the minimization of a cost function ${\cal C}({\mf \lambda})$ \cite{GrNaSo09}.

We demonstrate the concept and performance of nonlinear VQC by emulating it classically for the canonical example of the time-independent one-dimensional nonlinear Schr\"{o}dinger equation
\begin{align}
 \left[ -\frac{1}{2} \frac{\text{d}^{2}}{\text{d} x^{2}} + V(x) + g |f(x)|^{2} \right] f(x) = E f(x) \, , \label{gpe}
\end{align}
where $V$ is an external potential and $g$ denotes the strength of the nonlinearity.
We also implement nonlinear VQC for Eq.~\eqref{gpe} on IBM quantum computers to establish its feasibility on current NISQ devices.
Proof-of-principle results are shown in Figs.~\ref{fig:1}(c) and \ref{fig:1}(d) demonstrating excellent agreement with numerically exact solutions.
The nonlinear Schr\"{o}dinger equation and its generalizations to higher dimensions describe various physical phenomena ranging from Bose-Einstein condensation to light propagation in nonlinear media~\cite{Gr61, Pi61, Le01, PiSt03, Sc05, Ag13}.
In particular, below we consider Eq.~\eqref{gpe} with quasi-periodic potentials $V$ that are in the focus of current cold atom experiments~\cite{ViEtAl19} and that make Eq.~\eqref{gpe} challenging to solve numerically.
The methods used for this equation here are straightforwardly modified to handle other nonlinear terms and time-dependent problems as illustrated in \cite{SupplementalMaterial} for the Burgers equation appearing in fluid dynamics.

The ground state of Eq.~\eqref{gpe} can be found by minimizing the cost function
\begin{align}
 \mc{C} = \mcl{K}_{c} + \mcl{P}_{c} + \mcl{I}_{c} \, , \label{cost}
\end{align}
where $\mcl{K}_{c}$, $\mcl{P}_{c}$ and $\mcl{I}_{c}$ are the mean kinetic, potential and interaction energies, respectively.
In Eq.~\eqref{cost} $\mcl{\cdot}_{c}$ denotes averages with respect to a single real-valued function $f^{(1)} \equiv f$ on the interval $[a, b]$ satisfying the normalization condition $\int_{a}^{b} |f(x)|^{2} \text{d}x = 1$.

In line with standard numerical approaches~\cite{PrEtAl92, GrRoSt07, Is09} we apply the finite difference method (FDM) to Eq.~\eqref{gpe} and discretize the interval $[a, b]$ into $N$ equidistant grid points $x_{k} = a + h_{N} k$, where $h_{N} = \len / N$ is the grid spacing, $\len = b-a$ is the length of the interval and $k \in \{0, \ldots, N-1\}$.
Each grid point is associated with a variational parameter $f_{k}$ that approximates the continuous solution $f(x_{k})$ at $x_{k}$.
Furthermore, we impose periodic boundary conditions (i.e., $f_{N} = f_{0}$) and the normalization condition imposed on the continuous functions $f$ translates to
\begin{align}
 1 = h_{N} \sum_{k = 0}^{N-1} |f_{k}|^{2} = \sum_{k = 0}^{N-1} |\psi_{k}|^{2} \, , \label{norm}
\end{align}
where $\psi_{k} = \sqrt{h_{N}} f_{k}$.
Note that the condition on the set of parameters $\{ \psi_{k} \}$ is independent of the grid spacing, and in the following we consider optimizing the cost function with respect to them.

All averages $\mcl{\cdot}_{c}$  in Eq.~\eqref{cost} can be approximated by corresponding expressions of the discrete problem $\mcl{\cdot}$.
We find~\cite{PrEtAl92, GrRoSt07, Is09} $\mcl{\cdot}_{c} = \mcl{\cdot} + \mc{E}_{\text{grid}}$, where $\mc{E}_{\text{grid}} \propto 1 / N^{2}$ is the error associated with the trapezoidal rule when transforming integrals into sums, and
\begin{subequations}
\label{grid}
\begin{align}
 \mcl{K} & = -\frac{1}{2} \frac{1}{h_{N}^{2}} \sum_{k = 0}^{N-1} \psi_{k}^{*} \left( \psi_{k+1} - 2\psi_{k} + \psi_{k-1} \right) \, , \label{kin} \\
 \mcl{P} & = \sum_{k = 0}^{N-1} \left[ \psi_{k}^{*} V(x_{k}) \psi_{k} \right] \, , \label{pot} \\
 \mcl{I} & = \frac{1}{2} \frac{g}{h_{N}} \sum_{k = 0}^{N-1} |\psi_{k}|^{4} \, . \label{int}
\end{align}
\end{subequations}
Note that $\mcl{K}$ in Eq.~\eqref{kin} uses a FDM representation of the second-order derivative in Eq.~\eqref{gpe}.

For evaluating the terms in Eq.~\eqref{grid} on a quantum computer we consider quantum registers with $n$ qubits and basis states $\ket{\mf q}=\ket{q_{1} \ldots q_{n}} = \ket{q_{1}} \otimes \ldots \otimes \ket{q_{n}}$, where $q_{j} \in \{0, 1\}$ denotes the computational states of qubit $j$.
Regarding the sequence $q_{1} \ldots q_{n} = \text{binary}(k)$ as the binary representation of the integer $k = \sum_{j = 1}^{n} q_{j} 2^{n-j}$, we encode all $N = 2^{n}$ amplitudes $\psi_{k}$ in the normalized state
\begin{align}
 \ket{\psi} = \sum_{k = 0}^{N-1} \psi_{k} \ket{\text{binary}(k)} \, . \label{quantum}
\end{align}
We prepare the quantum register in a variational state $\ket{\psi({\mf \lambda})}$ via the quantum circuit $\hat{U}({\mf \lambda})$ of depth $d$ shown in Fig.~\ref{fig:1}(b).
We consider depths $d \propto \text{poly}(n)$ such that the quantum ansatz requires exponentially fewer parameters $\mc{N}$ than standard classical schemes with $N$ parameters.
Note that the number of variational parameters scales like $\mc{N} \propto n d$ for our quantum ansatz~\cite{SupplementalMaterial}.
The power of this ansatz is rooted in the fact that it encompasses all matrix product states (MPS)~\cite{VeMuCi08, Os10, Os11, Or14} with bond dimension $\chi \sim \text{poly}(n)$~\cite{SupplementalMaterial}.
Since polynomials and Fourier series~\cite{Kh11, Os13} can be efficiently represented by MPS, the quantum ansatz simultaneously contains universal basis functions that are capable of approximating a large class of solutions to nonlinear problems efficiently.
Furthermore, we show below that the quantum ansatz is capable of storing solutions with exponentially fewer resources than the classically optimized MPS ansatz.
Note that the number of variational parameters scales like $\mc{N} \propto n \chi^{2}$ for MPS of bond dimension $\chi$~\cite{SupplementalMaterial}.

\begin{figure}[t!]
\centering
\includegraphics[width=86.387mm]{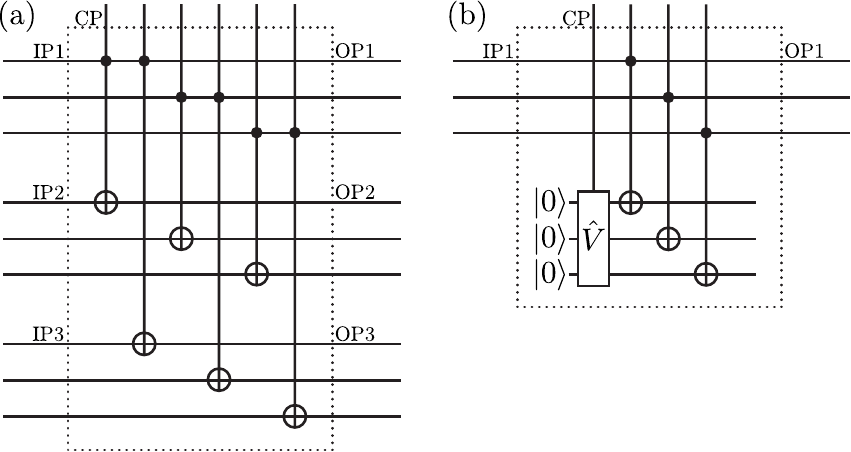}
\caption{\label{fig:2}
(a) QNPU circuit calculating the nonlinear term $|\psi|^{4}$.
The networks are shown for $n = 3$ and all input ports IP are fed the same variational quantum states created by $\hat{U}({\mf \lambda})$.
IP2 is fed $\hat{U}({\mf \lambda})$ and IP3 is fed $\hat{U}^{*}({\mf \lambda})$.
(b) QNPU circuit for working out the potential energy term $\tilde{V} |\psi|^{2}$.
}
\end{figure}

Figure~\ref{fig:2}(a) demonstrates the basic working principle of the QNPU for the nonlinear term $\mcl{I}$.
The effect of the controlled NOT operations between pairs of qubits is to provide a point-wise multiplication with the ancilla thus measuring $\sum_{k} |\psi_{k}|^{4}$.
In Fig.~\ref{fig:2}(b) we show the circuit for measuring $\mcl{P}$.
The unitary $\hat{V} \hat{=} \, O_{1}$ encodes function values of the external potential $V$.
A copy of $\psi$ is effectively multiplied point-wise with the external potential by controlled NOT gates to give $\sum_{k} \tilde{V}_{k} |\psi_k|^2$.
Similarly, multiplying $\psi$ with their shifted versions using adder circuits (see \cite{SupplementalMaterial} for details) allows evaluating the kinetic energy term.

The measured expectation value of the ancilla qubit is directly related to the desired quantities as $\mcl{I} = g \ma{\hat{\sigma}_{z}}^{I} / 2 h_{N}$ for the nonlinear term, $\mcl{P} = \alpha \ma{\hat{\sigma}_{z}}^{P}$ for the potential energy and $\mcl{K} = \left( 1 - \ma{\hat{\sigma}_{z}}^{K} \right)/h_{N}^{2}$ for the kinetic energy~\cite{EkEtAl02, AlEtAl03}.
Furthermore, derivatives of the cost function, as required by some minimization algorithms \cite{Sp92, GrNaSo09, LuMoJa18}, can be evaluated by combining the ideas presented here with the quantum circuits discussed in~\cite{LiBe17, McEtAl18, OBEtAl19, MiNaMi19}.

The unitary network $\hat{V}$ represents scaled function values $\tilde{V}_{k}$ of the external potential where $\sum_{k = 0}^{N-1} |\tilde{V}_{k}|^{2} = 1$, and $\alpha > 0$ a scaling parameter such that $V_{k} = \alpha \tilde{V}_{k}$.
Efficient quantum circuits $\hat{V}$ for measuring $\mcl{P}$ can be systematically obtained by establishing tensor networks as a programming paradigm.
To this end we expand the external potential in polynomials or Fourier series, $\tilde{V}(x) \approx \sum_{j}^{J} c_{j} b_{j}(x)$, where $b_{j}(x)$ are basis functions and $c_{j}$ are expansion coefficients~\cite{SupplementalMaterial}.
In the case of Fourier series of order $J$, the approximate potential is represented by an MPS of bond dimension $\chi = J$~\cite{Kh11,Os13}.
Next we write the MPS in terms of $n - \lceil\log\chi\rceil$ unitaries~\cite{ScEtAl05, ScEtAl07, BaEtAl08}, where $\lceil\cdot\rceil$ is the ceiling function.
Each of these unitaries acts on  $2\chi$ qubits and can be decomposed in terms of elementary two-qubit gates~\cite{ReEtAl94, BaEtAl95, ShBuMa06, ItEtAl16, SupplementalMaterial}.
An upper bound for the depth of the resulting quantum circuit is $d_{>} \le 9n[(23/48)(2\chi)^2 + 4/3]$~\cite{ShBuMa06}.
The depth thus scales polynomially with the number of qubits $n$ and with $\chi$, and many problems of interest show an even more advantageous scaling.
For example, in the following we consider the  potential
\begin{align}
 V(x) = s_{1} \sin(\kappa_{1} x) + s_{2} \sin(\kappa_{2} x) \label{disV}
\end{align}
and set $\kappa_{2} = 2 \kappa_{1} / (1 + \sqrt{5})$.
This potential realizes an incommensurate bichromatic lattice where the ratio $s_{1} / s_{2}$ determines the amount of disorder in the lattice~\cite{DeEtAl10}.
The trap potential $V(x)$ in Eq.~\eqref{disV} is exactly represented by an MPS of bond dimension $\chi = 4$.
The depth of the corresponding quantum circuit $d = 5(n-2)+1 \ll d_{>}$ is much smaller than the upper bound~\cite{SupplementalMaterial}.

Next we analyze the Monte Carlo sampling error~\cite{PrEtAl92} associated with the measurement of the ancilla qubit.
We denote the absolute sampling error associated with quantity $X$ by $\mc{E}_{\text{MC}}^{X}$, and the corresponding relative error is~\cite{SupplementalMaterial}
\begin{subequations}
\label{error}
\begin{align}
 \mc{\epsilon}_{\text{MC}}^{P} & = \frac{\mc{E}_{\text{MC}}^{P}}{\mcl{P}} = C_{P} \frac{1}{\sqrt{M}} \, , \label{erP} \\
 \mc{\epsilon}_{\text{MC}}^{K} & = \frac{\mc{E}_{\text{MC}}^{K}}{\mcl{K}} \approx C_{K} \frac{N}{N_{\text{min}}} \frac{1}{\sqrt{M}} \, , \label{erK} \\
 \mc{\epsilon}_{\text{MC}}^{I} & = \frac{\mc{E}_{\text{MC}}^{I}}{\mcl{I}} \approx C_{I} \frac{N}{N_{\text{min}}} \frac{1}{\sqrt{M}} \, . \label{erI}
\end{align}
\end{subequations}
In this equation, we assume  $N \ge N_{\text{min}}$ and $N_{\text{min}} = \len / \len_{\text{min}}$  is the minimal number of grid points for resolving the smallest length scale $\len_{\text{min}}$ of the problem.
The parameters $C_{X}$ in Eq.~\eqref{error} are of the order of unity~\cite{SupplementalMaterial}, and all sampling errors decrease with the number of samples $M$ as $1 / \sqrt{M}$.
While the relative error associated with the potential term in Eq.~\eqref{erP} is independent of the number of grid points, $\mc{\epsilon}_{\text{MC}}^{K}$ and $\mc{\epsilon}_{\text{MC}}^{I}$ increase linearly with $N/N_{\text{min}}$.
It follows that increasing the grid size requires larger values of $M$ in order to keep the sampling error small.
However, the grid error scales like $\mc{E}_{\text{grid}} \propto 1 / N^{2} = 2^{-2n}$ for $N \ge N_{\text{min}}$~\cite{SupplementalMaterial}.
We thus conclude that only moderate ratios $N / N_{\text{min}} > 1$ and therefore relatively small values of $M$ are needed in order to achieve accurate solutions with small grid errors.

\begin{figure}[t!]
\centering
\includegraphics[width=86.42mm]{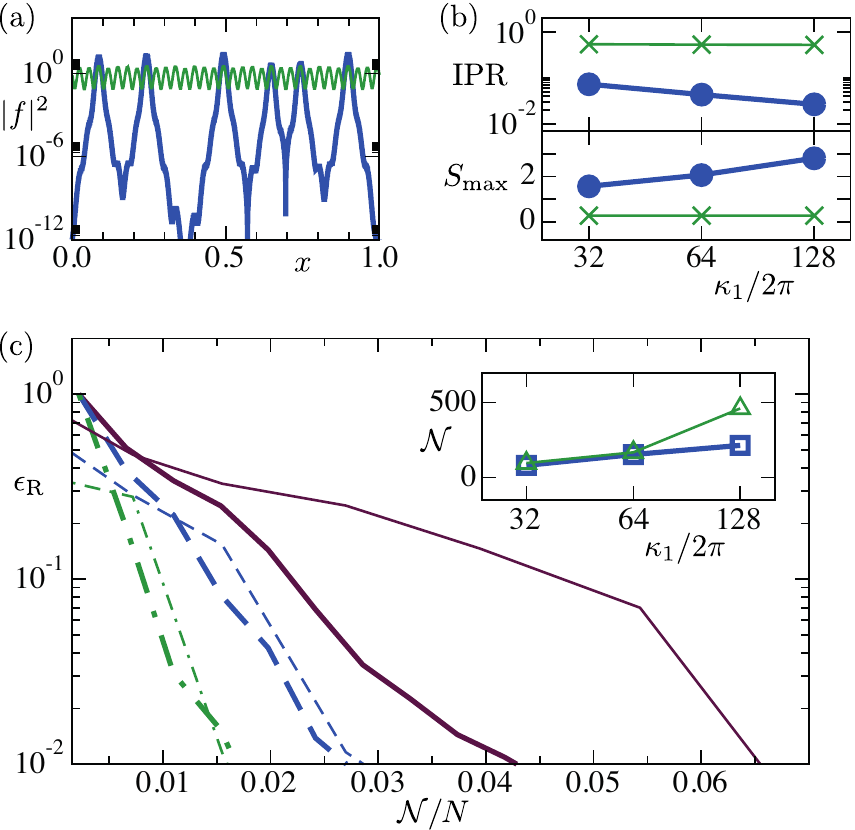}
\caption{\label{fig:3}
(a) Numerically exact solution $|f(x)|^{2}$ of Eq.~\eqref{gpe} on a logarithmic scale, $V(x)$ in Eq.~\eqref{disV} with $s_{1} = 2 \times 10^{4}$ and $\kappa_{1} = 2 \pi \times 32$.
The green thin line (blue normal line) is for  $s_{1} / s_{2} = 200$ ($s_{1} / s_{2} = 2$).
(b) Log-log plot of the IPR (top panel) and lin-log plot of $S_{\text{max}}$ of the exact solution $\ket{\psi^{\text{exact}}}$ (bottom panel) as a function of $\kappa_1$.
Green crosses (blue dots) correspond to $s_{1} / s_{2} = 200$ ($s_{1} / s_{2} = 2$).
(c) Representation error $\epsilon_{\text{R}}$ of the exact solution for the quantum ansatz (thick lines) and the MPS ansatz (thin lines) as a function of $\mc{N} / N$.
All curves are for $s_{1} / s_{2} = 2$ and correspond to $s_{1} = 2 \times 10^{4}$ and $\kappa_{1} = 2 \pi \times 32$ (green dash-dotted), $s_{1} = 8 \times 10^{4}$ and $\kappa_{1} = 2 \pi \times 64$ (blue dashed), and $s_{1} = 3.2 \times 10^{5}$ and $\kappa_{1} = 2 \pi \times 128$ (purple solid), respectively.
The inset shows $\mc{N}$ as a function of $\kappa_{1}$ for $\epsilon_{\text{R}} = 0.05$ and $s_{1} / s_{2} = 2$.
Blue squares (green triangles) correspond to the quantum ansatz (MPS ansatz).
All curves in (a)-(c) are for $N = 2^{13} = 8192$ grid points and $g = 50$.
}
\end{figure}

The quantum ansatz in Fig.~\ref{fig:1}(b) is inspired from tensor network theory and can be regarded, for example, as the Trotter decomposition of the time $t$ evolution operator $\exp(-\imag H(t) t/\hbar)$ of a time-dependent spin Hamiltonian $H(t)$ with arbitrary short-range interactions acting on the initial state $\ket{\mf 0}$~\cite{DaEtAl04}.
Similarly to the coupled cluster ansatz in quantum chemistry VQC calculations~\cite{PeEtAl14}, there is currently no known efficient classical ansatz for this state~\cite{TrEtAl12, ScEtAl13}.
From the VQC perspective, our quantum ansatz in Fig.~\ref{fig:1}(b) is composed of generic two-qubit gates that in an experiment are accurately approximated by short sequences of gates if a sufficiently tunable or universal gate set is experimentally available~\cite{NiCh10}.
We envisage that this quantum ansatz is more efficient than methods based on an MPS ansatz on a classical computer like the multigrid renormalization (MGR) method in~\cite{LuMoJa18}.
This MGR method is the most efficient and accurate classical algorithm known to us for the problem considered here, for which it can already be exponentially faster than standard classical algorithms.
A comparison with this powerful classical method, which is based on variational classical MPS, will allow us to validate the superior variational power of the quantum ansatz in Fig.~\ref{fig:1}(b) on a quantum computer.
Note that the difficulty of solving a classical optimization problem in many variables does not go away by using the quantum ansatz, as the actual optimization is classical and there is no quantum advantage there.
The quantum advantage stems solely from the faster evaluation of the cost function for our quantum ansatz in Fig.~\ref{fig:1}(b) on a quantum computer.

To provide numerical evidence for this we first obtain the numerically exact solution of Eq.~\eqref{gpe} on the interval $[0, 1]$ via the MGR algorithm~\cite{LuMoJa18} and by allowing for the maximal bond dimension $\chi$ of the MPS ansatz~\cite{SupplementalMaterial}.
In this case the numerically exact solution is described by $N = 2^{n}$ parameters like in other conventional algorithms.
The results are shown in Fig.~\ref{fig:3} (a) for two different values of $s_{1} / s_{2}$.
In the weakly disordered regime $s_{1} / s_{2} \gg 1$, $|f(x)|^{2}$ varies on the length scale set by $1 / \kappa_{1}$.
On the contrary, the strongly disordered regime $s_{1} / s_{2} \approx 1$ is characterized by strongly localized solutions in space.
The localization of the wavefunction can be quantified using the inverse participation ratio (IPR)~\cite{KrMa93}, $\text{IPR} = (N \sum_{k = 0}^{N-1} |\psi_{k}|^{4})^{-1}$.
We show the IPR in Fig.~\ref{fig:3}(b) as a function of $\kappa_1$ (top panel) and find that it stays constant for $s_{1} / s_{2} \gg 1$.
On the other hand, the IPR decreases according to a power law with $\kappa_1$ for $s_{1} / s_{2} \approx 1$, showing that the localized character of the wavefunction increases dramatically with $\kappa_1$.

Next we encode the function values of the numerically exact solution in the state $\ket{\psi^{\mathrm{exact}}}$ via Eq.~\eqref{quantum} and calculate the maximum bipartite entanglement entropy $S_{\text{max}}$ of all possible bipartitions of
the $n$-qubit wave function $\ket{\psi^{\mathrm{exact}}}$.
The quantity $S_{\text{max}}$ is a measure of the entanglement of $\ket{\psi^{\mathrm{exact}}}$ and is shown in the bottom panel of Fig.~\ref{fig:3}(b).
The value of $S_{\text{max}}$ is small and stays constant with $\kappa_{1}$ for $s_{1}/s_{2} \gg 1$ in the weakly disordered regime.
Contrary to this, for $s_{1}/s_{2} \approx 1$ in the strongly disordered regime we observe that $S_{\text{max}} \propto \log(\kappa_{1})$.

The entanglement measure $S_{\text{max}}$ provides a useful necessary criterion for efficient MPS approximations~\cite{Vi03}.
A MPS of bond dimension $\chi$ can at most contain an amount of entanglement $S_{\text{max}}[\text{MPS}] \leq \log(\chi)$~\cite{VeMuCi08, Or14}.
For MPS to be efficient, we require that $\chi$ scales at most polynomially with $n$, i.e.\ $\chi = \text{poly}(n)$, so that $S_{\text{max}}[\text{MPS}] \leq \log(\text{poly}(n))$ and therefore MPS can only capture small amounts of entanglement efficiently.
The small values of $S_{\text{max}}$ for $s_{1}/s_{2} \gg 1$ suggest that MPS work well in the weakly disordered regime and indeed we have confirmed numerically that this is true.
Therefore, in the following we focus on the strongly disordered regime $s_{1}/s_{2} \approx 1$.
In this regime MPS cannot be an efficient approximation for large values of $\kappa_{1}$:
The total number of variational parameters of MPS $\mc{N}[\text{MPS}]$ depends quadratically on $\chi$ [35], i.e.\ $\mc{N}[\text{MPS}] \propto \chi^{2}$, and $\chi$ needs to grow polynomially with $\kappa_{1}$, i.e.\ $\chi \propto \text{poly}(\kappa_{1})$, to satisfy the observed entanglement scaling, such that $\mc{N}[\text{MPS}] \propto \text{poly}(\kappa_{1})$.
Our quantum ansatz (QA) of Fig.~\ref{fig:1}(b) can capture much larger amounts of entanglement $S_{\text{max}}[\text{QA}] \lesssim d$ efficiently~\cite{ScEtAl13} and therefore this ansatz can be an efficient approximation for large values of $\kappa_{1}$:
Because the total number of variational parameters $\mc{N}$ depends linearly on $d$ for the quantum ansatz~\cite{SupplementalMaterial}, i.e.\ $\mc{N}[\text{QA}] \propto d$, and $d$ just needs to grow logarithmically, i.e.\ $d \propto \log(\kappa_{1})$, for the observed entanglement requirements, we conclude that $\mc{N}[\text{QA}] \propto \log(\kappa_{1})$.
These entanglement considerations show that the quantum ansatz has the potential to be exponentially more efficient than MPS in the strongly disordered regime for increasing values of $\kappa_{1}$.

To quantitatively analyze and demonstrate the efficiency of the quantum ansatz in this regime, we obtain the set of parameters ${\mf \lambda}$ that maximize the fidelity $\mc{F} = | \braket{\psi^{\mathrm{exact}}}{\psi({\mf \lambda})} |$  for different depths $d$~\cite{SupplementalMaterial}.
The infidelity $\epsilon_{\text{R}} = 1 - \mc{F}$ is thus a measure of the error when approximating the exact solution by this ansatz, and in the following we refer to $\epsilon_{\text{R}}$ as the representation error.
As shown in Fig.~\ref{fig:3}(c), the representation error decreases exponentially as a function of $\mc{N}$ for all values of $\kappa_{1}$ and therefore we obtain accurate solutions for $\mc{N} / N \ll 1$.
Even for the largest value of $\kappa_{1} = 2 \pi \times 128$ and $\epsilon_{\text{R}} \approx 10^{-2}$, we find $\mc{N} / N \approx 0.04$ so that we only require $4 \%$ of the full number of parameters needed in conventional algorithms.
Most importantly, the inset of Fig.~\ref{fig:3}(c) shows that, to obtain a fixed representation error of $\epsilon_{\text{R}} = 0.05$, the number of parameters of our quantum ansatz needs to grow as $\mc{N}[\text{QA}] \propto \log(\kappa_{1})$.
This numerical analysis therefore confirms our expectation, from the entanglement arguments in the previous paragraph, that the quantum ansatz efficiently approximates solutions in the strongly disordered regime even for large values of $\kappa_{1}$.
This is possible because the quantum ansatz captures the required entanglement $S_{\text{max}} \propto \log(\kappa_{1})$ by means of just the small number of parameters $\mc{N}[\text{QA}] \propto \log(\kappa_{1})$.
The entanglement capabilities of MPS imply that $\mc{N}[\text{MPS}] \propto \mathrm{poly}(\kappa_{1})$ has to be fulfilled.
Therefore we conclude that our quantum ansatz is exponentially more efficient than MPS in the strongly disordered regime for growing values of $\kappa_{1}$.
This key finding shows that nonlinear VQC can be exponentially more efficient than optimized, classical variational schemes that are based on the MPS ansatz.

We now propose a calculation that illustrates the exponential advantage of our quantum ansatz on a quantum computer particularly clearly.
We consider the ground state problem of Eq.~\eqref{gpe} with the quasi-random external potential of Eq.~\eqref{disV} on the interval $[0, 1]$ discretized by $N = 2^{n}$ equidistant grid points and use the FDM from above that leads to Eqs.~\eqref{kin}, \eqref{pot}, and \eqref{int}.
We choose $\kappa_{1} = 2 \pi \times 2^{n/2}$ and assume that $n = 4, 6, 8, \ldots$ qubits store the variational quantum ansatz.
The smallest wave length present in our external potential of Eq.~\eqref{disV} is determined by $\kappa_{1}$ and reads $2 \pi / \kappa_{1} = 2^{-n/2}$.
The interval $[0, 1]$ accomodates $1 / 2^{-n/2} = 2^{n/2}$ of such wave lengths and our grid of $N = 2^{n}$ equidistant points resolves each such wave length using $N / 2^{n/2} = 2^{n/2}$ grid points.
Therefore the randomness of our quasi-random external potential of Eq.~\eqref{disV} as well as its FDM resolution grow exponentially with the number of qubits $n$.
At the same time, all FDM errors (using a FDM representation of the Laplace operator, the trapezoidal rule for integration, and so on) decrease polynomially with the grid spacing $1 / N = 2^{-n}$ and thus exponentially with $n$.
Therefore, with growing values of $n$, our FDM representation of Eq.~\eqref{gpe} converges exponentially fast to the continuous problem and the accuracy of randomness in Eq.~\eqref{disV} grows exponentially.
Our proposal is to solve this problem for the strongly disordered regime $s_{1} / s_{2} \approx 1$ in Eq.~\eqref{disV}, i.e.\ compute the corresponding ground states, using an increasing number $n = 4, 6, 8, \ldots$.
Based on the numerical analysis of Fig.~\ref{fig:3}, we conjecture that MPS require resources $\mc{N}[\text{MPS}] \propto \text{poly}(\kappa_{1}) = \text{poly}(2 \pi \times 2^{n/2}) = \text{exp}(n)$ growing exponentially with $n$, whereas the quantum ansatz of Fig.~\ref{fig:1}(b) just needs resources $\mc{N}[\text{QA}] \propto \log(\kappa_{1}) = \log(2 \pi \times 2^{n/2}) = \text{poly}(n)$ increasing polynomially with $n$.
This calculation can be used to demonstrate quantum supremacy as, by successively increasing $n$, the limit of what is computationally possible on a classical computer (using MPS or exact classical methods) is reached quickly and our quantum ansatz on a quantum computer remains efficient far beyond this classical limit.

The quantum hardware requirements for this quantum supremacy calculation go beyond the current capabilities of available NISQ devices~\cite{CrEtAl19}.
Nevertheless the superior performance of our quantum ansatz is relevant for current NISQ devices as only the most efficient variational states can succeed in the presence of the current experimental errors.
We test the feasibility of nonlinear VQC on NISQ devices by calculating the ground state of Eq.~\eqref{gpe} for a simple harmonic potential and a single variational parameter on an IBM Q device~\cite{ibmq} utilizing further network optimizations (see \cite{SupplementalMaterial} for details).
The experimental implementation of the nonlinear VQC algorithm was able to identify the optimal variational parameter with an error of less than $10\%$ leading to excellent agreement of the ground state solutions with exact numerical solutions [c.f.~Figs.~\ref{fig:1}(c)(d)].

The methods presented here are readily modified to two-and three-dimensional problems with an overhead scaling linearly in the number of dimensions, and can be applied to a broad range of nonlinear terms and differential operators.
An exciting prospect for future work will be to utilize intermediate-scale quantum computers for solving non-linear problems on grid sizes beyond the scope of conventional computers.

\begin{acknowledgments}
ML and DJ are grateful for funding from the Networked Quantum Information Technologies Hub (NQIT) of the UK National Quantum Technology Programme as well as from the EPSRC grant ``Tensor Network Theory for strongly correlated quantum systems'' (EP/K038311/1).
We acknowledge support from the EPSRC National Quantum Technology Hub in Networked Quantum Information Technology (EP/M013243/1).
MK and DJ acknowledge financial support from the National Research Foundation and the Ministry of Education, Singapore.
\end{acknowledgments}

\bibliography{bibliography}

\end{document}


\title{Supplemental Material:\\Variational quantum algorithms for nonlinear problems}

\author{Michael Lubasch${}^1$}

\author{Jaewoo Joo${}^1$}

\author{Pierre Moinier${}^2$}

\author{Martin Kiffner${}^{3, 1}$}

\author{Dieter Jaksch${}^{1, 3}$}

\affiliation{Clarendon Laboratory, University of Oxford, Parks Road, Oxford OX1 3PU, United Kingdom${}^1$}
\affiliation{BAE Systems, Computational Engineering, Buckingham House, FPC 267 PO Box 5, Filton, Bristol BS34 7QW, United Kingdom${}^2$}
\affiliation{Centre for Quantum Technologies, National University of Singapore, 3 Science Drive 2, Singapore 117543${}^3$}

\maketitle

First we provide the details for the IBM quantum computer experiments in Sec.~\ref{sec:1}.
Then Sec.~\ref{sec:2} contains a description of the adder circuit used in the main text for computing the kinetic energy.
Section~\ref{sec:3} presents how to do time evolution with the Burgers equation using the variational framework.
We explain the matrix product state (MPS) ansatz in Sec.~\ref{sec:4} and derive its quantum circuit representation.
The Monte Carlo sampling error is explained in Sec.~\ref{sec:5}.
In Sec.~\ref{sec:6} we describe the fidelity optimization algorithm and provide the number of variational parameters in the quantum and MPS ansatz.
We use the same notation and definitions as in the main text.

\section{IBM Quantum Computations}
\label{sec:1}

\begin{figure}
\centering
\includegraphics[width=86.397mm]{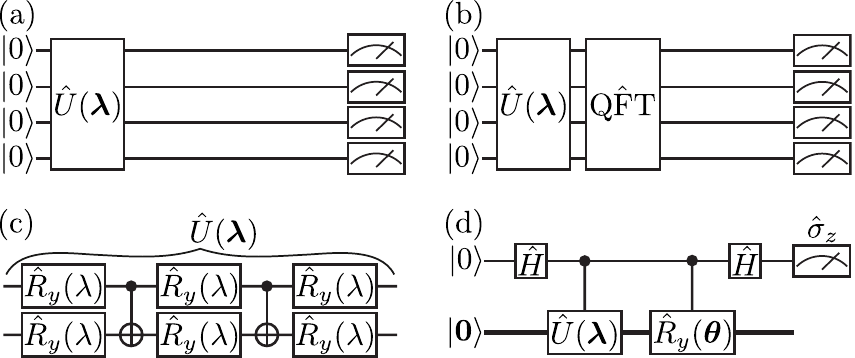}
\caption{\label{fig:supp1}
(a) Quantum circuits for expectation value computation of the potential $\langle \psi | \hat{D}_{\tilde{V}} | \psi \rangle = \sum_{k} |\psi_{k}|^{2} \tilde{V}_{k}$ or nonlinear term $\langle \psi | \hat{D}_{|\psi|^{2}} | \psi \rangle = \sum_{k} |\psi_{k}|^{2} |\psi_{k}|^{2}$ in the nonlinear Schr\"{o}dinger equation, where $|\psi\rangle = \hat{U}(\boldsymbol{\lambda}) |\boldsymbol{0}\rangle$.
(b) Quantum circuit for expectation value computation of the Laplace operator $\langle \psi | \hat{\Delta} | \psi \rangle$.
(c) The quantum ansatz that we have used as variational state on the IBM quantum computer.
Here $\hat{R}_{y}(\lambda) = \exp(-\imag \lambda \hat{\sigma}_{y} / 2)$ is the standard $y$ rotation gate.
The quantum ansatz has just one variational parameter, i.e.\ $\boldsymbol{\lambda} = \lambda$.
In the IBM experiment we are restricting $\lambda$ to be a real parameter and therefore the quantum ansatz has only real entries.
(d) Quantum circuit for evaluating specific or averaged function values $\Re \{ \psi_{k} \}$ where $\Re \{ \cdot \}$ denotes the real part.
The gates $\hat{R}_{y}(\boldsymbol{\theta})$ are individually controlled rotations around the $y$-axis and depending on $\boldsymbol{\theta}$ allow reading out function values and averages of $\Re \{ \psi_{k} \}$ by measuring the ancilla qubit in the computational basis~\cite{EkEtAl02, AlEtAl03}.
For example, individual function values can be measured efficiently at an arbitrary grid point $k$ by choosing rotation angles $\boldsymbol{\theta} = \pi \, \text{binary}(k)$.
Averages over the least significant qubits can be formed by choosing the corresponding rotation angles $\pi / 2$ when reading the function on a coarse grid.
We obtain the imaginary part $\Im \{ \psi_{k} \}$ by including an additional phase shift gate $R_{-\pi/2} = \text{diag}(1, \exp(-\imag \pi/2))$ directly before the second Hadamard gate.
Function values of more general functions $F = \prod_{j=1}^{r} (O_{j} f^{(j)})$ are evaluated by combining the circuit presented here with the QNPU from Fig.~1 in the main text.
}
\end{figure}

To run on IBM's noisy intermediate-term quantum computers, we need to reduce the hardware requirements and optimize the circuits shown in the main text.
These optimizations lead to deviations from the generic nonlinear VQC circuit structure discussed in the main text.
For computing the expectation value of an operator $\hat{O}$ that is diagonal, $\hat{O} = \hat{D}_{O}$, or can be diagonalized easily, i.e.\ $\hat{O} = \hat{D}_{O} = \sum_{k} o_{k} |k \rangle \langle k|$, we use a quantum circuit that does not require ancilla qubits.
The expectation value of such an operator reads $\langle \psi | \hat{D}_{O} | \psi \rangle = \sum_{k} o_{k} \langle \psi | k \rangle \langle k | \psi \rangle = \sum_{k} | \langle k | \psi \rangle |^{2} o_{k}$.
This expectation value is obtained by measuring all qubits and computing the mean value after $M$ experiments as $\langle \psi | \hat{D}_{O} | \psi \rangle \approx (1/M) \sum_{m = 1}^{M} o_{k^{(m)}}$, as shown in Fig.~\ref{fig:supp1} (a).
The following hardware-reduced quantum algorithms were originally inspired by the standard methods that were used in previous experiments for the experimental computation of expectation values of quantum chemistry Hamiltonians, e.g.~\cite{PeEtAl14, KaEtAl17}.

Each measurement $m$ gives a multi-index $\text{binary}(k)^{(m)}$ from which we obtain $k^{(m)}$.
Then the expectation values read $\langle \psi | \hat{D}_{\tilde{V}} | \psi \rangle \approx (1/M) \sum_{m = 1}^{M} \tilde{V}_{k^{(m)}}$ and $\langle \psi | \hat{D}_{|\psi|^{2}} | \psi \rangle \approx (1/M) \sum_{m = 1}^{M} |\psi_{k^{(m)}}|^{2}$.
We compute $|\psi_{k}|^{2}$ for a particular value of $k$ by counting how often that value of $k$ appears in $M$ experiments and then dividing that number by $M$.

Figure~\ref{fig:supp1} (b) shows the corresponding quantum circuit for the Laplace operator $\Delta$.
The Laplace operator is diagonalized by the Quantum Fourier Transform (QFT) and has eigenvalues $\Delta_{k} = 2 N^{2} \left( \cos(2 \pi k / N) - 1 \right)$ where $N = 2^{n}$ for $n$ qubits.
Therefore $\langle \psi | \hat{\Delta} | \psi \rangle = \langle \psi | (\hat{\text{QFT}})^{\dag} \hat{D}_{\Delta} (\hat{\text{QFT}}) | \psi \rangle =  \sum_{k} |\psi_{k}|^{2} \Delta_{k}$ where $\psi_{k}$ is obtained from the QFT of $| \psi \rangle$, i.e.\ $(\hat{\text{QFT}}) | \psi \rangle$.
After we have applied the QFT to $| \psi \rangle = \hat{U}(\boldsymbol{\lambda}) |\boldsymbol{0}\rangle$, we compute the expectation value as in Fig.~\ref{fig:supp1} (a).

To illustrate that this quantum algorithm works, we chose the smallest non-trivial system size of $n = 2$ qubits and the variational ansatz of Fig.~\ref{fig:supp1} (c) which is parametrized by one real parameter $\lambda$.
To make use of the best available hardware, we chose to work on the 20 qubit IBM devices Tokyo and Poughkeepsie.
On the Tokyo device, the best T1 is $148.5 \mu s$, the best T2 is $78.4 \mu s$, the best two-qubit error rate is $1.47 \%$, and the best single-qubit error rate is $0.064 \%$~\cite{IBMSpecs}.
On the Poughkeepsie device, the best T1 is $123.3 \mu s$, the best T2 is $123.6 \mu s$, the best two-qubit error rate is $1.11 \%$, and the best single-qubit error rate is $0.052 \%$~\cite{IBMSpecs}.
These experimental errors are just a little bit too large for our quantum algorithm presented in the main text but are good enough for our hardware-efficient quantum algorithm presented here (see also~\cite{CrEtAl19}).
We considered the nonlinear Schr\"{o}dinger equation with a harmonic trap potential $V(x) = 2000(x-0.5)^{2}$ and $x \in [0, 1)$.
Using the variational ansatz, we evaluated the quantum circuits of Fig.~\ref{fig:supp1} (a) and (b) for all values of $\lambda \in [0, 2\pi)$ in steps of $0.1$.
For each value of $\lambda$ we computed the average of the particular quantity of interest over 24000 experimental shots.
This allowed us to compute the cost function (total energy in the nonlinear Schr\"{o}dinger equation) of Fig.~1 in the main text and determine its minimum as well as the corresponding solution function.
Fig.~1 in the main text shows the squared absolute value of the solution functions, which is also referred to as the ground state solution (density).
This is compared to the numerically exact solution obtained from imaginary time evolution~\cite{LiYn98}.

Figure~\ref{fig:supp1} (d) shows the general quantum circuit for evaluating specific or averaged function values.
We did not use this circuit in the IBM experiment where, instead, we used the circuit of Fig.~\ref{fig:supp1} (a) to compute $|\psi_{0}|^{2}$, $|\psi_{1}|^{2}$, $|\psi_{2}|^{2}$, and $|\psi_{3}|^{2}$.

Note that the QFT required here could be realized using just single-qubit gates with classical control~\cite{GrNi96}, which would simplify the hardware requirements significantly.
We did not exploit this simplification here since the IBM quantum computers do not offer classical control yet.
Therefore we used the complete QFT quantum circuit shown in Fig.~\ref{fig:supp1} (b).

\section{Adder Circuit for Kinetic Energy}
\label{sec:2}

\begin{figure}
\centering
\includegraphics[width=62.529mm]{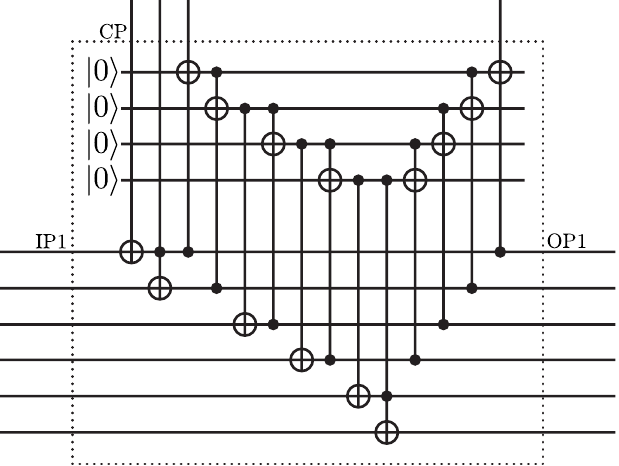}
\caption{\label{fig:supp2}
QNPU for computing the kinetic energy in the main text via the adder operator $\hat{A}$.
This operator acts on a state $|\psi\rangle = \sum_{k=0}^{N-1} \psi_{k} |\text{binary}(k)\rangle$ in such a way that $\hat{A} \sum_{k=0}^{N-1} \psi_{k} |\text{binary}(k)\rangle = \sum_{k=0}^{N-1} \psi_{k+1} |\text{binary}(k)\rangle$.
We consider periodic boundary conditions where $\psi_{N} = \psi_{0}$.
Inserting this QNPU in Fig.~1 (a) in the main text gives the circuit for determining the kinetic energy $\llangle K \rrangle$.
Note that $\hat{U}(\boldsymbol{\lambda})$ is fed on input port IP1 and in $|\psi\rangle = \hat{U}(\boldsymbol{\lambda}) |\boldsymbol{0}\rangle$, the lowermost qubit is the most significant qubit and the uppermost qubit is the least significant.
}
\end{figure}

The quantum circuit presented in Fig.~\ref{fig:supp2} is the QNPU for computing the kinetic energy in the main text via the adder operator $\hat{A}$.
This operator increments the index of every wave function coefficient by one and can therefore be seen as a special case of the general quantum circuits for arithmetic operations discussed in Refs.~\cite{VeBaEk96, NiCh10}.
Compared with the more general circuits of Refs.~\cite{VeBaEk96, NiCh10}, this adder circuit requires fewer ancilla qubits and gates.
We deduce from Fig.~\ref{fig:supp2} that for $n > 2$ qubits, the adder circuit is composed of $n-2$ ancilla qubits, $n-2$ CNOT and $2n-2$ Toffoli gates.
For $n = 2$ the adder circuit requires one CNOT and one Toffoli gate, and for $n = 1$ one CNOT gate suffices.

\section{Nonlinear VQC for the Burgers Equation}
\label{sec:3}

To illustrate how a time-dependent nonlinear partial differential equation is solved, we develop a quantum algorithm that approximates time evolution of the Burgers equation
\begin{align}\label{eq:BE}
 \frac{\partial}{\partial t} f(x, t) & = \nu \frac{\partial^{2}}{\partial x^{2}} f(x, t) - f(x, t) \frac{\partial}{\partial x} f(x, t) \, .
\end{align}
Here, $\nu$ is the so-called coefficient of kinematic viscosity and Eq.~\eqref{eq:BE} gives rise to turbulence when $1/\nu$ becomes large~\cite{BeKh07, Wh11}.
We discretize the spatial coordinate $x$ (as done in the main text of this article for the nonlinear Schr\"{o}dinger equation) and we represent the resulting function values $f_{k}$ at time $t$ by a vector $|f\rangle$.
Then Eq.~\eqref{eq:BE} takes on the form:
\begin{align}\label{eq:BE2}
 \frac{\partial}{\partial t} |f\rangle & = \nu \Delta |f\rangle - D_{f} \nabla |f\rangle \, ,
\end{align}
where $\Delta$ and $\nabla$ are the discretized Laplace and Nabla matrix, respectively, and $D_{f}$ is a diagonal matrix with the values $f_{k}$ on its diagonal.

For the classical simulation of Eq.~\eqref{eq:BE2}, we use the time-dependent variational principle algorithm for MPS~\cite{LuOsVa15, HaEtAl16}.
This algorithm requires that $\Delta$ and $|f\rangle \nabla$ are written as matrix product operators and we show how this is done in Ref.~\cite{LuMoJa18}.

For the quantum algorithm, it is important to note that Eq.~\eqref{eq:BE2} does not conserve the norm of $|f\rangle$.
Therefore we choose the variational ansatz to be $|f\rangle = \lambda_{0} |\psi(\boldsymbol{\lambda})\rangle$, where $\lambda_{0}$ is a new variational parameter and $\boldsymbol{\lambda} = (\lambda_{1}, \lambda_{2}, \ldots)$ as before (see discussion in main text).
The introduction of $\lambda_{0}$ allows us to handle arbitrary norms $\langle f | f \rangle$ of the variational solution $|f\rangle$, while the wave function on the quantum computer always fulfills $\langle \psi(\boldsymbol{\lambda}) | \psi(\boldsymbol{\lambda}) \rangle = 1$.

To keep this discussion as simple as possible, we just consider the Euler method for time evolution~\cite{PrEtAl92} here.
The Euler method computes the time-evolved solution after time step $\tau$, $|f(t+\tau)\rangle$, by using the previous solution, $|f(t)\rangle$, and the right-hand side of Eq.~\ref{eq:BE2} at time $t$.
If we summarize the right-hand side of Eq.~\ref{eq:BE2} by $O |f\rangle$, then the Euler method identifies $|f(t+\tau)\rangle = \left( \mathds{1} + \tau O(t) \right) |f(t)\rangle$.
In the variational setting, instead of directly computing $|f(t+\tau)\rangle$ via this formula, it is more efficient to define the cost function
\begin{align}\label{eq:BE3}
 \mathcal{C}(|f(t+\tau)\rangle) & = || |f(t+\tau)\rangle - (\mathds{1} + \tau O(t)) |f(t)\rangle||^{2}
\end{align}
and minimize this cost function via the variational parameters of $|f(t+\tau)\rangle$.

\begin{figure}
\centering
\includegraphics[width=86.376mm]{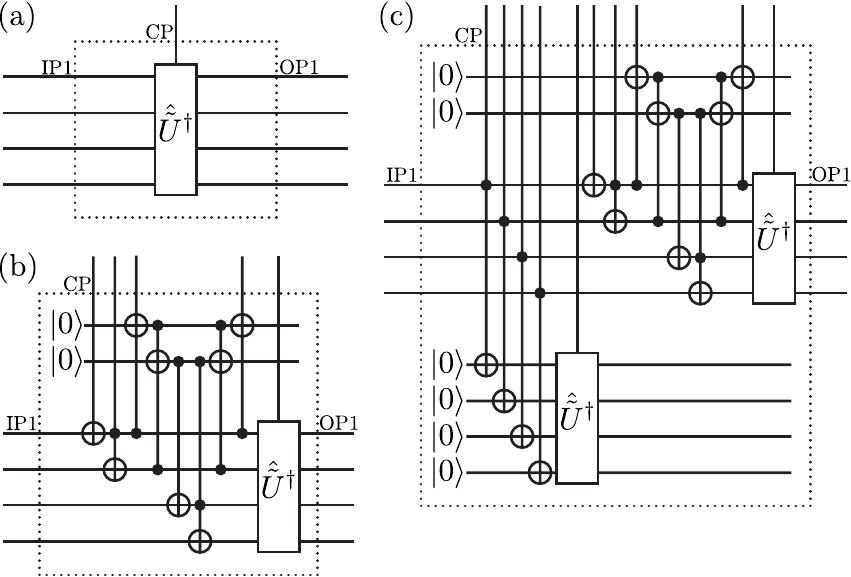}
\caption{\label{fig:supp3}
QNPUs for evaluating the cost function Eq.~\eqref{eq:BE4} of the Burgers equation for an example with $n = 4$.
These QNPUs need to be inserted in the circuit of Fig.~1 (a) from the main text with a small modification: in this circuit the unitary on input port IP1 needs to be controlled by the ancilla (as are all unitaries on the other input ports IP2 to IP$r$).
(a) $\langle \tilde{\psi} | \psi \rangle = \langle \boldsymbol{0} | \hat{\tilde{U}}^{\dag} \hat{U}(\boldsymbol{\lambda}) | \boldsymbol{0} \rangle$.
(b) $\langle \tilde{\psi} | \hat{A} | \psi \rangle = \langle \boldsymbol{0} | \hat{\tilde{U}}^{\dag} \hat{A} \hat{U}(\boldsymbol{\lambda}) | \boldsymbol{0} \rangle$.
(c) $\langle \tilde{\psi} | \hat{A} \hat{D}_{\tilde{\psi}}^{\dag} | \psi \rangle = \langle \boldsymbol{0} | \hat{\tilde{U}}^{\dag} \hat{A} \hat{D}_{\tilde{\psi}}^{\dag} \hat{U}(\boldsymbol{\lambda}) | \boldsymbol{0} \rangle$.
Here $\hat{\tilde{U}}^{\dag}$ is the adjoint of $\hat{\tilde{U}}$ which defines $|\tilde{\psi}\rangle = \hat{\tilde{U}} |\boldsymbol{0}\rangle$, i.e.\ the function $|f(t)\rangle = \tilde{\lambda}_{0} |\tilde{\psi}\rangle$ at time $t$.
Note that $\hat{\tilde{U}}^{\dag}$ is fixed and has no variational parameters.
The variational parameters reside outside the QNPUs in $\hat{U}(\boldsymbol{\lambda})$ which defines $|f(t+\tau)\rangle = \lambda_{0} |\psi\rangle = \lambda_{0} \hat{U}(\boldsymbol{\lambda}) |\boldsymbol{0}\rangle$.
Furthermore, in the notation here, $\hat{A}$ is the adder circuit defined in the caption of Fig.~\ref{fig:supp2}, and $\hat{D}_{\tilde{\psi}}$ denotes the diagonal operator that has the values of $\tilde{\psi}_{k}$ on its diagonal.
}
\end{figure}

For the quantum algorithm we define $|f(t+\tau)\rangle = \lambda_{0} |\psi(\boldsymbol{\lambda})\rangle = \lambda_{0} \hat{U}(\boldsymbol{\lambda}) |\boldsymbol{0}\rangle$ and $|f(t)\rangle = \tilde{\lambda}_{0} |\tilde{\psi}\rangle = \tilde{\lambda}_{0} \hat{\tilde{U}} |\boldsymbol{0}\rangle$ for every value of $t$.
Then for each time step $\tau$, the following cost function needs to be minimized:
\begin{align}\label{eq:BE4}
 \mathcal{C}(\lambda_{0}, \boldsymbol{\lambda}) & = ||\lambda_{0} |\psi(\boldsymbol{\lambda})\rangle - (\mathds{1}+\tau \hat{O}) \tilde{\lambda}_{0} |\tilde{\psi}\rangle||^{2} \\
                                                                              & = |\lambda_{0}|^{2} - 2 \Re \{ \lambda_{0} \tilde{\lambda}_{0}^{*} \langle \tilde{\psi} | (\mathds{1}+\tau \hat{O}) | \psi(\boldsymbol{\lambda}) \rangle \} + \mathrm{const.} \nonumber \\
                                                                              & = |\lambda_{0}|^{2} - 2 \Re \{ \lambda_{0} \tilde{\lambda}_{0}^{*} \langle \boldsymbol{0} | \hat{\tilde{U}}^{\dag} (\mathds{1}+\tau \hat{O}) \hat{U}(\boldsymbol{\lambda}) | \boldsymbol{0} \rangle \} + \mathrm{const.} , \nonumber
\end{align}
where $\hat{O} = \nu \Delta - \tilde{\lambda}_{0} \hat{D}_{\tilde{\psi}} \nabla$.
For each time step, Eq.~\eqref{eq:BE4} is minimized via the procedures discussed in Refs.~\cite{KaEtAl17, KiSw17}.
Figure~\ref{fig:supp3} shows the quantum circuits that are required for the computation of this cost function.
Note that the QNPUs need to be modified after every time step, as $\hat{\tilde{U}}$ changes after every time step.

More sophisticated time evolution algorithms can readily be realized for nonlinear partial differential equations on a quantum computer.
For example, a time-dependent variational principle can be formulated similarly to Refs.~\cite{LiBe17, McEtAl18} where the required new ingredients (for nonlinear terms) are obtained from the quantum circuits presented here and in the main text.

\section{MPS Ansatz}
\label{sec:4}

In this section, we describe the general procedure for transforming MPS into quantum circuits.
We also discuss how plane wave, cosine, and sine functions are represented on a quantum computer.

\subsection{General Procedure}

An MPS of bond dimension $\chi$,
\begin{align}\label{eq:MPS}
 |\psi^{\text{MPS}}\rangle & = \sum_{q_{1}, \ldots, q_{n}} B[1]^{q_{1}} \ldots B[n]^{q_{n}} | q_{1}, \ldots, q_{n} \rangle \, ,
\end{align}
is defined in terms of $n$ tensors $B[j]_{\alpha_{j-1}, \alpha_{j}}^{q_{j}}$, where the index $q_{j} \in \{0, 1 \}$ and all indices $\alpha_{j}$ run from $1$ to $\chi$ except $\alpha_{0}$ and $\alpha_{n}$ which both take on only the value $1$.
The MPS tensor entries are real or complex numbers and constitute the variational parameters in the MPS ansatz.
We observe that there are $\mathcal{O}(n \chi^{2}) =\mathcal{O}(\text{poly}(n))$ variational parameters, bipartite entanglement entropy is limited to values $\lesssim \log(\chi)$, and most MPS algorithms have a computational cost $\mathcal{O}(\text{poly}(\chi))$~\cite{VeMuCi08, Or14}.

\begin{figure}
\centering
\includegraphics[width=85.896mm]{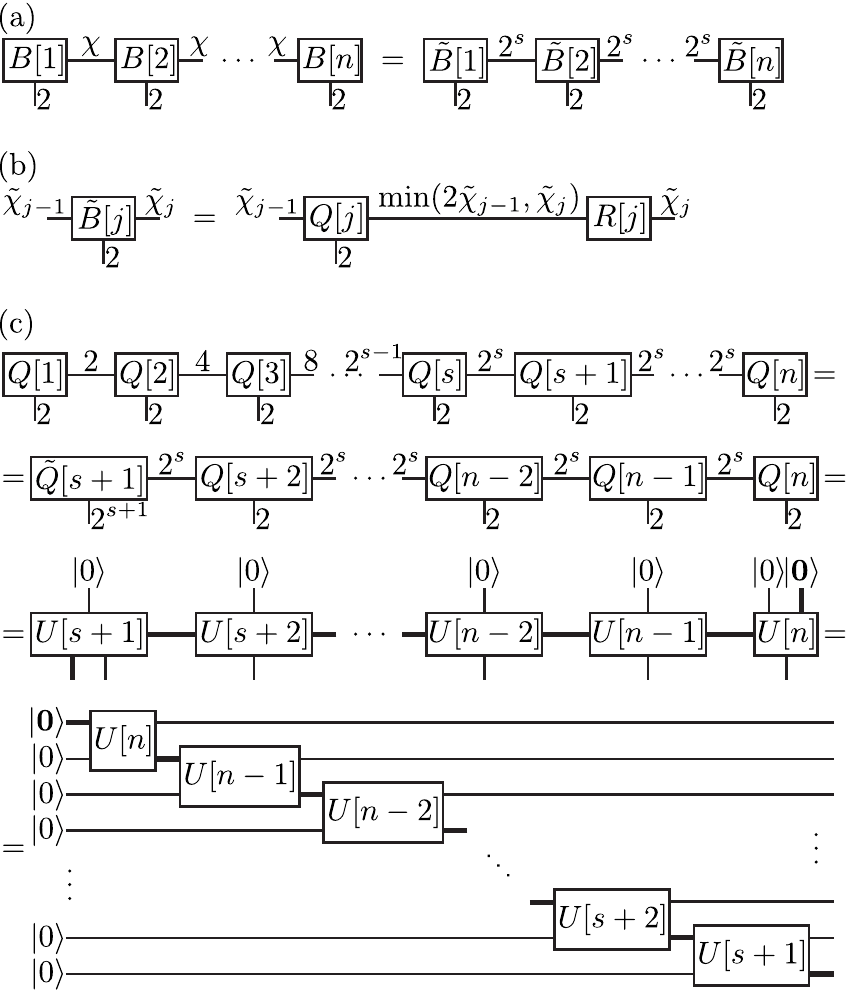}
\caption{\label{fig:supp4}
Transformation from MPS into quantum circuit.
(a) We first compute the value $s = \lceil \log_{2}(\chi) \rceil$, where $\lceil \cdot \rceil$ is the ceiling function.
This allows us to rewrite the MPS with bond dimension $\tilde{\chi} = 2^{s}$.
If $\tilde{\chi} > \chi$ then we achieve this by embedding each tensor $B[j]$ in a new larger tensor $\tilde{B}[j]$ where the additional new elements are set to zero.
(b) Then successive QR decompositions turn this MPS into a product of isometric matrices (and possibly a normalization factor).
We start with the QR decomposition of $\tilde{B}[1]$ and move towards and end with $\tilde{B}[n]$.
After the QR decomposition of $B[j]$ we multiply $R[j]$ with $\tilde{B}[j+1]$ before performing the QR decomposition of $\tilde{B}[j+1]$.
The resulting tensor network consists of isometric matrices only.
(c) We multiply all tensors left of $Q[s+1]$ with $Q[s+1]$ to obtain $\tilde{Q}[s+1]$.
The resulting isometric matrices are then embedded in unitary matrices $U[j]$.
Rearranging the final network gives the MPS quantum circuit.
}
\end{figure}

\begin{figure}
\centering
\includegraphics[width=85.815mm]{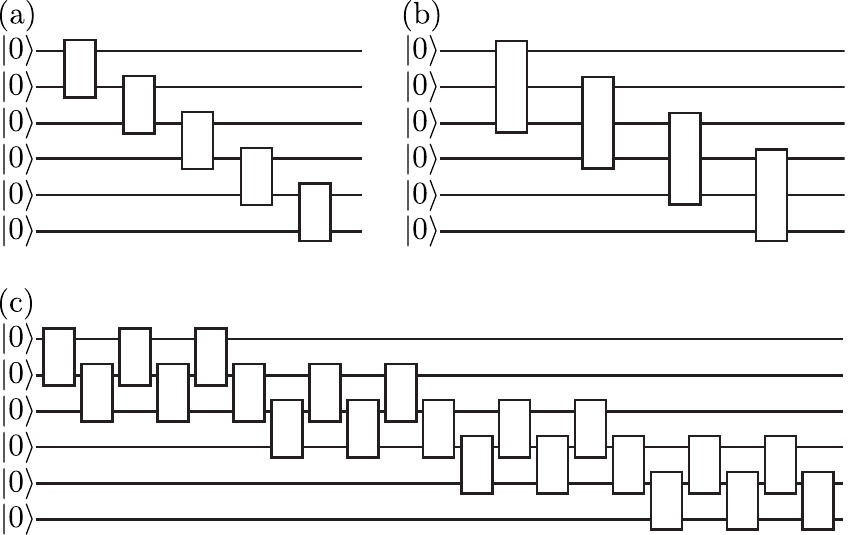}
\caption{\label{fig:supp5}
Example MPS quantum circuits.
(a) MPS of bond dimension $\chi = 2$ represented in terms of $n-1$ generic two-qubit unitaries.
(b) MPS of bond dimension $\chi = 4$ represented in terms of $n-2$ generic three-qubit unitaries.
(c) MPS of bond dimension $\chi = 4$ represented in terms of $5 (n-2) + 1$ generic two-qubit unitaries using the decomposition of three-qubit unitaries of Ref.~\cite{BaEtAl95}.
}
\end{figure}

Figure~\ref{fig:supp4} shows the required steps for transforming a MPS of bond dimension $\chi$ into a quantum circuit.
The MPS quantum circuit consists of $n-s$ unitary matrices each of dimension $2^{s+1} \times 2^{s+1} = 2\tilde{\chi} \times 2\tilde{\chi}$.
Each of these unitary matrices can be further decomposed into a product of $\mathcal{O}(\tilde{\chi}^{2})$ generic two-qubit unitaries~\cite{ShBuMa06}.
We can embed these resulting two-qubit unitaries in the quantum ansatz of Fig.~1 (b) in the main text.
Therefore any MPS of bond dimension $\chi \sim \text{poly}(n)$ is efficiently contained in the quantum ansatz of depth $d \sim \text{poly}(n)$.
Figure~\ref{fig:supp5} illustrates how the resulting quantum circuits look like for an MPS of bond dimensions $\chi = 2$ and $4$.
For the depths $d$ of these quantum circuits we obtain $d = n-1$ for $\chi = 2$ and $d = 5 (n-2) + 1$ for $\chi = 4$.

\subsection{Plane Wave, Sine, and Cosine Function Representations on a Quantum Computer}

\begin{figure}
\centering
\includegraphics[width=86.12mm]{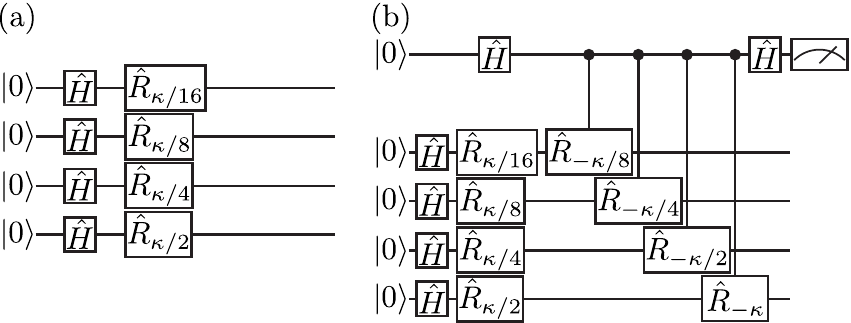}
\caption{\label{fig:supp6}
(a) This quantum circuit realizes the wave function $|\psi\rangle = (1/\sqrt{2^{n}}) \sum_{q_{1}, q_{2}, \ldots, q_{n}} \exp(\text{i} \kappa \sum_{j=1}^{n} q_{j} 2^{-j}) |q_{1}, q_{2}, \ldots, q_{n}\rangle$ which represents a plane wave of wave vector $\kappa$.
$R_{\varphi} = \text{diag}(1, \exp(\text{i} \varphi))$ is a phase shift gate of phase shift $\varphi$.
(b) Quantum circuit for the probabilistic generation of $\psi(x) \propto \cos(\kappa x)$ and $\psi(x) \propto \sin(\kappa x)$ with probability $0.5$, respectively.
When $0$ is measured in the uppermost qubit, then $|\psi\rangle = (1/\sqrt{2^{n}}) \sum_{q_{1}, q_{2}, \ldots, q_{n}} \exp(\text{i} \kappa \sum_{j=1}^{n} q_{j} 2^{-j}) + \exp(-\text{i} \kappa \sum_{j=1}^{n} q_{j} 2^{-j}) |q_{1}, q_{2}, \ldots, q_{n}\rangle$, and the wave function coefficients are proportional to the cosine function.
When $1$ is measured in the uppermost qubit, then $|\psi\rangle = (1/\sqrt{2^{n}}) \sum_{q_{1}, q_{2}, \ldots, q_{n}} \exp(\text{i} \kappa \sum_{j=1}^{n} q_{j} 2^{-j}) - \exp(-\text{i} \kappa \sum_{j=1}^{n} q_{j} 2^{-j}) |q_{1}, q_{2}, \ldots, q_{n}\rangle$, and the wave function coefficients are proportional to the sine function.
}
\end{figure}

MPS contain polynomials as well as Fourier series~\cite{Kh11, Os13} and so these are also contained in the quantum ansatz presented in Fig.~1 (b) in the main text.
A plane wave function $\exp(\text{i} \kappa x)$ can be written as an MPS of bond dimension $\chi = 1$~\cite{Kh11} and its quantum circuit representation requires only single-qubit gates, as can be seen in Fig.~\ref{fig:supp6} (a).
We write the function $\sin(\kappa x)$ as an MPS of bond dimension $\chi = 2$~\cite{Kh11}, which is a quantum circuit of the form shown in Fig.~\ref{fig:supp5} (a).
We write the sum of two different sine functions as an MPS of bond dimension $\chi = 4$, which is a quantum circuit of the form shown in Fig.~\ref{fig:supp5} (b) and (c).
This MPS of bond dimension $\chi = 4$ is used in the main text for representing the potential $V(x) = s_{1} \sin(\kappa_{1} x) + s_{2} \sin(\kappa_{2} x)$.

Depending on the hardware capabilities, there exist several alternatives for realizing $V(x) = s_{1} \sin(\kappa_{1} x) + s_{2} \sin(\kappa_{2} x)$ on a quantum computer.
Note that $\sin(\kappa x)$ as well as $\cos(\kappa x)$ can be written as a sum of two plane waves, respectively.
Therefore two quantum circuits of the form of Fig.~\ref{fig:supp6} (a) suffice to represent a single function $\sin(\kappa x)$.
Four quantum circuits of the form of Fig.~\ref{fig:supp6} (a) represent the potential $V(x) = s_{1} \sin(\kappa_{1} x) + s_{2} \sin(\kappa_{2} x)$.
Therefore, instead of a single MPS quantum circuit, we can also use several simpler circuits (composed of single-qubit unitaries only) for the potential representation on a quantum computer.
Sine and cosine functions are also generated probabilistically with the quantum circuit of Fig.~\ref{fig:supp6} (b).

\section{Sampling Error}
\label{sec:5}

Here we analyze the grid error and the Monte Carlo sampling error.
We first investigate the convergence of the numerically exact solution with the number of grid points.
These results facilitate the analysis of the Monte Carlo sampling error.

\subsection{Grid Error}

We numerically calculate the expectation values $\mcl{K}$, $\mcl{P}$ and $\mcl{I}$ in the main text for different grid sizes $N = 2^{n}$.
The results are presented in Fig.~\ref{fig:supp7} and show that all expectation values increase with an increasing number of grid points until they level off and converge at a critical number of grid points $N_{\text{min}} = 2^{n_{\text{min}}}$.
This critical number depends only on the value of the wavenumber $\kappa_{1}$ which controls the shape of the external potential.
More specifically, $N_{\text{min}}$ approximately coincides with the number of grid points required to resolve the potential
\begin{align}
 V(x) & = s_{1} \sin(\kappa_{1} x) + s_{2} \sin(\kappa_{2} x)\,.
\end{align}
By taking into account that one needs at least four grid points to resolve a single wave length, we obtain $n_{\text{min}} = 6$ for $\kappa_{1} = 2 \pi \times 16$, $n_{\text{min}} = 7$ for $\kappa_{1} = 2 \pi \times 32$, $n_{\text{min}} = 8$ for $\kappa_{1} = 2 \pi \times 64$, and $n_{\text{min}} = 9$ for $\kappa_{1} = 2 \pi \times 128$.
These simple estimates are in very good agreement with the results in Fig.~\ref{fig:supp7}.
More generally, we define
\begin{align}\label{lenscale}
 N_{\text{min}} = \frac{\len}{\len_{\text{min}}}\,,
\end{align}
where $\len_{\text{min}}$ is the smallest lengthscale of the problem.
The expectation values $\mcl{\cdot}$ converge if the grid size obeys $N > N_{\text{min}}$, and only in this regime the grid error scales like $\mc{E}_{\text{grid}} \propto 1 / N^{2}$.

The expectation values $\mcl{X}$ are related in the main text to expectation values of an ancilla qubit $\ma{\hat{\sigma}_{z}}^{X}$ when measured on a quantum computer.
The results for $\ma{\hat{\sigma}_{z}}^{X}$ are also shown in Fig.~\ref{fig:supp7} and illustrate how the values $\ma{\hat{\sigma}_{z}}^{X}$ converge for $N > N_{\text{min}}$.
We observe that $\ma{\hat{\sigma}_{z}}^{P}$ converges to a constant value and
\begin{subequations}\label{ndep}
 \begin{align}
 1 - \ma{\hat{\sigma}_{z}}^{K} \propto 2^{-2n}\,,\\
  \ma{\hat{\sigma}_{z}}^{I} \propto 2^{-n}\,,
 \end{align}
\end{subequations}
for $n > n_{\text{min}}$.
In particular, we thus conclude $\ma{\hat{\sigma}_{z}}^{K} \rightarrow 1$ and $\ma{\hat{\sigma}_{z}}^{I} \rightarrow 0$ for $n \rightarrow \infty$.
Note that the expected scaling in Eq.~\eqref{ndep} can be used in practice to determine the minimal grid size $N_{\text{min}}$.
This is important for finding the optimal grid size that results in converged results with reasonable Monte Carlo sampling sizes, as discussed in the next section.

\subsection{Monte Carlo Sampling Error}

The expectation values $\ma{\hat{\sigma}_{z}}$ in Eqs.~(6) and~(7) of the main text represent exact quantum mechanical expectation values.
Measuring these quantities on a quantum computer involves averaging over $M$ Monte Carlo samples,
\begin{align}
 \maov{\hat{\sigma}_{z}} = \frac{1}{M} \sum\limits_{m=1}^{M} \mq{\hat{\sigma}_{z}}_{m}\,,
\end{align}
where $\mq{\hat{\sigma}_{z}}_{m}$ denotes the eigenvalue of $\hat{\sigma}_{z}$ obtained for the $m$th sample.
In general, the Monte Carlo sampling error $\mc{E}_{\text{MC}} = \maov{\hat{\sigma}_{z}} - \ma{\hat{\sigma}_{z}}$ associated with drawing $M$ random samples is~\cite{PrEtAl92}
\begin{subequations}\label{MCE}
\begin{align}
 \mc{E}_{\text{MC}} & = \frac{\sqrt{\ma{\hat{\sigma}_{z}^{2}} - \ma{\hat{\sigma}_{z}}^{2}}}{\sqrt{M}}\,, \\
                                & = \frac{\sqrt{1 - \ma{\hat{\sigma}_{z}}^{2}}}{\sqrt{M}}\,,
\end{align}
\end{subequations}
where we used $\hat{\sigma}_{z}^{2} = \mathds{1}$ in the second line.

\begin{figure}
\centering
\includegraphics[width=86.186mm]{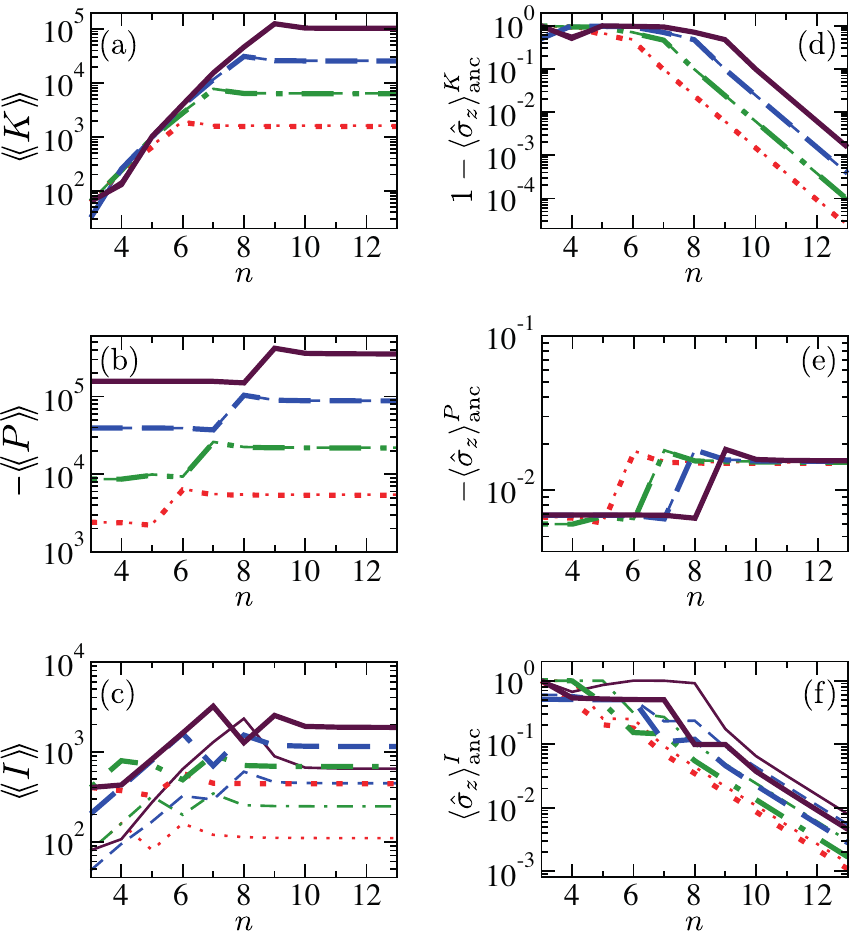}
\caption{\label{fig:supp7}
Energy expectation values $\mcl{K} = 4^{n} \Re \{ \langle \psi | (\mathds{1} - \hat{A}) | \psi \rangle \}$ (a), $\mcl{P} = \Re \{ \langle \psi | \hat{V} | \psi \rangle \}$ (b), $\mcl{I} = g 2^{n} \Re \{ \langle \psi | \hat{D}_{|\psi|^{2}} | \psi \rangle \}$ (c), and the corresponding values related to the ancilla qubit measurement:
$1 - \ma{\hat{\sigma}_{z}}^{K} = \mcl{K} / 4^{n}$ (d), $\ma{\hat{\sigma}_{z}}^{P} = \mcl{P} / \alpha$ (e), and $\ma{\hat{\sigma}_{z}}^{I} = \mcl{I} / (g 2^{n})$ (f).
Here $\hat{A}$ is the adder operator defined in the caption of Fig.~\ref{fig:supp2}, $\hat{V}$ represents the potential $V(x) = s_{1} \sin(\kappa_{1} x) + s_{2} \sin(\kappa_{2} x)$, and $\hat{D}_{|\psi|^{2}}$ is diagonal with the values of $|\psi_{k}|^{2}$ on its diagonal.
We consider the strongly disordered regime of Fig.~3 in the main text (where $s_{1}/s_{2} = 2$ and $\kappa_{2} = 2 \kappa_{1} / (1 + \sqrt{5})$) with $s_{1} = 5 \times 10^{3}$ and $\kappa_{1} = 2 \pi \times 16$ (red dotted), $s_{1} = 2 \times 10^{4}$ and $\kappa_{1} = 2 \pi \times 32$ (green dash-dotted), $s_{1} = 8 \times 10^{4}$ and $\kappa_{1} = 2 \pi \times 64$ (blue dashed), $s_{1} = 3.2 \times 10^{5}$ and $\kappa_{1} = 2 \pi \times 128$ (purple solid), and we compare $g = 10$ (thin lines) with $50$ (normal lines).
All results are for $N = 2^{13}$ grid points.
}
\end{figure}

With the help of these definitions we obtain
\begin{subequations}\label{MCerr}
\begin{align}
 \mc{E}_{\text{MC}}^{P} & = \alpha \frac{\sqrt{1 - [\ma{\hat{\sigma}_{z}}^{P}]^{2}}}{\sqrt{M}}\,, \\
 \mc{E}_{\text{MC}}^{K} & = \frac{N^{2}}{\ell^{2}} \frac{\sqrt{1 - [\ma{\hat{\sigma}_{z}}^{K}]^{2}}}{\sqrt{M}} \,,\\
 \mc{E}_{\text{MC}}^{I} & = \frac{1}{2} g \frac{N}{\len} \frac{\sqrt{1 - [\ma{\hat{\sigma}_{z}}^{I}]^{2}}}{\sqrt{M}} \,, \label{errI}
\end{align}
\end{subequations}
where $\mc{E}_{\text{MC}}^{X}$ is the absolute Monte Carlo error associated with the average $\mcl{X}$.
The expressions in Eq.~\eqref{MCerr} can be further simplified, and we begin with the error associated with the potential energy $\mcl{P}$.
With the help of Eq.~(7)(a) in the main text, the relative Monte Carlo sampling error can be written as
\begin{align}
 \mc{\epsilon}_{\text{MC}}^{P} & = \frac{\mc{E}_{\text{MC}}^{P}}{\mcl{P}} = C_{P} \frac{1}{\sqrt{M}}\,,
\end{align}
where
\begin{align}
 C_{P} = \sqrt{\frac{\alpha^2}{\mcl{P}^2} -1}\,.
\end{align}
Since $\alpha^{2}/\mcl{P}^{2} \ge 1$ by virtue of the definition of $\alpha$ in the main text, $C_{P}$ is always real.
For the parameters of Fig.~\ref{fig:supp7} $C_{P}$ takes on values between $64.3$ and $67.2$.

Next we investigate the error associated with the kinetic energy term $\mcl{K}$ and note that
\begin{subequations}
\begin{align}
 1 - [\ma{\hat{\sigma}_{z}}^{K}]^{2} & = \left(1 - \ma{\hat{\sigma}_{z}}^{K}\right) \left(1 + \ma{\hat{\sigma}_{z}}^{K}\right) \,, \\
 1 - \ma{\hat{\sigma}_{z}}^{K} & = h_{N}^{2} \mcl{K}\,. \label{K2}
\end{align}
\end{subequations}

Combining the equations above we obtain the relative sampling error associated with $\mcl{K}$,
\begin{align}
 \mc{\epsilon}_{\text{MC}}^{K} & = \frac{ \mc{E}_{\text{MC}}^{K}}{ \mcl{K}} = \frac{N}{\len} \frac{\sqrt{1 + \ma{\hat{\sigma}_{z}}^{K}}}{\sqrt{ \mcl{K}}} \frac{1}{\sqrt{M} }\,. \label{errorSr}
\end{align}
To further simplify this we assume that $N \ge N_{\text{min}}$ such that $\ma{\hat{\sigma}_{z}}^{K} \approx 1$  [see Eq.~\eqref{ndep}] and utilize Eq.~\eqref{lenscale},
\begin{align}
 \mc{\epsilon}_{\text{MC}}^{K} \approx C_{K} \frac{N}{N_{\text{min}}} \frac{1}{\sqrt{M}}\,,
\end{align}
where
\begin{align}
 C_{K} = \frac{\sqrt{2}}{l_{\text{min}}\sqrt{\mcl{K}}}\,.
\end{align}
The constant $C_{K}$ is expected to be of the order of unity since $\mcl{K}\approx 1/\len_{\text{min}}^{2}$.
For the parameters of Fig.~\ref{fig:supp7} $C_{K}$ takes on values between $2.24$ and $2.30$.

Finally, we consider the nonlinear term and point out that $\ma{\hat{\sigma}_{z}}^{I} \rightarrow 0$ for $N \rightarrow \infty$, see Eq.~\eqref{ndep}.
The leading term in Eq.~\eqref{errI} is thus $\mc{E}_{\text{MC}}^{I}\approx g N / (2\len\sqrt{M})$, and the relative Monte Carlo sampling error is
\begin{align}
 \mc{\epsilon}_{\text{MC}}^{I} & = \frac{\mc{E}_{\text{MC}}^{I}}{\mcl{I}} \approx C_{I} \frac{N}{N_{\text{min}}} \frac{1}{\sqrt{M}}\,,
\end{align}
where
\begin{align}
 C_{I} \approx \frac{g N_{\text{min}}}{2 \len \mcl{I}}\,.
\end{align}
For the parameters of Fig.~\ref{fig:supp7} $C_{I}$ takes on values between $2.58$ and $6.88$.

The preceeding analysis concludes the proof of Eq.~(8) in the main text.
For $N < N_{\text{min}}$ all observables require of the order of $M \approx \mc{\epsilon}_{\text{MC}}^{-2}$ Monte Carlo samples in order to be accurately measured.
The same result holds for the potential term even if $N \ge N_{\text{min}}$.
On the other hand, $\mcl{K}$ and $\mcl{I}$ require $M \approx 4^{n} \mc{\epsilon}_{\text{MC}}^{-2}$ experiments and thus $M$ increases exponentially with $n$.
The grid error also tends to zero exponentially fast $\propto 4^{-n}$ for $n > n_{\text{min}}$, and thus $n$ only needs to be slightly larger than $n_{\text{min}}$ in order to achieve accurate solutions.
This avoids exponentially many Monte Carlo samples.

Note that the minimal grid size $N_{\text{min}}$ can be determined by gradually increasing $N$ until the measured expectation values $\ma{\hat{\sigma}_{z}}$ show the expected scaling behaviour in Eq.~\eqref{ndep}.
In this way optimal grid sizes resulting in accurate solutions with moderate Monte Carlo sampling sizes can be chosen.

\section{Fidelity $\mc{F}$ Optimization and Number of Variational Parameters $\mathcal{N}$}
\label{sec:6}

In the main text of this article we determine optimal approximations in terms of MPS $|\psi^{\text{MPS}}\rangle$ and in terms of our quantum ansatz $|\psi^{\text{QA}}\rangle$ for numerically exact solutions $|\psi^{\text{exact}}\rangle$.
We achieve this by maximizing $\mc{F} = \langle \psi^{\text{exact}} | \psi^{\text{MPS}} \rangle$ as well as $\mc{F} = \langle \psi^{\text{exact}} | \psi^{\text{QA}} \rangle$ using standard tensor network optimization techniques~\cite{VeMuCi08, Or14}.
Our algorithm optimizes the variational unitaries one after another and for each unitary $U_{j}$ determines the new unitary $\tilde{U}_{j}$ that maximizes $\mc{F}$ under the assumption that all other unitaries are fixed.
Because $\mc{F} = \text{tr}(U_{j} R_{j})$ we obtain the optimal $\tilde{U}_{j}$ from a singular value decomposition of $R_{j} = U_{R_{j}} \Sigma_{R_{j}} V_{R_{j}}$ as $\tilde{U}_{j} = V_{R_{j}}^{\dag} U_{R_{j}}^{\dag}$.

For real-valued unitary gates (as considered throughout our analysis), our quantum ansatz of Fig.~1 (b) in the main text has 3 variational parameters per two-qubit unitary in the first column and then 6 variational parameters per two-qubit unitary in all the other columns.
A MPS tensor $B[j]_{\alpha_{j-1}, \alpha_{j}}^{q_{j}}$ has $\text{dim}(\alpha_{j-1}) \text{dim}(\alpha_{j}) \text{dim}(q_{j}) - \text{dim}(\alpha_{j})(\text{dim}(\alpha_{j})+1)/2$ variational parameters.
This is derived by reshaping $B[j]_{\alpha_{j-1}, \alpha_{j}}^{q_{j}} = B[j]_{(\alpha_{j-1}, q_{j}), \alpha_{j}}$ as an isometric matrix with row-muliindex $(\alpha_{j-1}, q_{j})$ and column-index $\alpha_{j}$.
This reshaped form allows us to count the number of free parameters of such an isometry.

\bibliography{bibliography}